\documentstyle[epsf,prc,aps]{revtex}
\begin{document}
\draft
\title{Mixed quark-nucleon phase in neutron stars and nuclear symmetry energy}
\author{M. Kutschera$^{1,2}$ and J. Niemiec$^1$}
\address{$^1$~H. Niewodnicza\'nski Institute of Nuclear Physics,
ul. Radzikowskiego 152, 31-342 Krak\'ow, Poland}
\address{$^2$~Institute of Physics, Jagiellonian University, ul. Reymonta 4, 
30-059 Krak\'ow, Poland }
\date{\today}
\maketitle
\begin{abstract}
The influence of the nuclear symmetry energy on the formation
of a~mixed quark-nucleon phase in neutron star cores is 
studied. We use simple parametrizations of the nuclear matter
equation of state, and the bag model for the quark phase. The
behavior of nucleon matter isobars, which is responsible for the
existence of the mixed phase, is investigated. 
The role of the nuclear symmetry energy changes 
with the value of the bag constant $B$. For lower values of $B$ the
properties of the mixed phase do not depend strongly on the
symmetry energy. For larger $B$ we find that a critical pressure for
the first quark droplets to form in the nucleon medium is strongly
dependent on the nuclear symmetry energy, but the pressure at which
last nucleons
disappear is independent of it. We
study the implications of these results for the structure of neutron
stars. The finite-size effects are also considered. We find
that the allowed range of  surface tension for the mixed
phase to be energetically favorable depends strongly on the
nuclear symmetry energy.
\end{abstract}

\section{Introduction}
Recently, Glendenning \cite{glen} has shown that a proper construction
of the nucleon-quark phase transition inside neutron stars
implies the coexistence of  nucleon matter and  quark matter
over a finite  range of pressure.
This has the effect that a core, or a spherical shell,
of a mixed quark-nucleon phase can exist inside  neutron stars. The fraction of
space occupied by quark matter 
smoothly increases from zero at  the core boundary, which
corresponds to a critical pressure for the first quark droplets
to form in the neutron star matter, to unity when eventually
the last nucleons dissolve into quarks. Heiselberg {\em et al.} \cite{heis} included
surface and Coulomb effects also in the mixed phase construction
and concluded that the mixed phase remains the ground state of the
neutron star matter for a physically reasonable range of surface tension.

In the original construction due to Glendenning, nucleon
matter was treated in the relativistic mean field (RMF) model.
The nuclear symmetry energy in the RMF model 
increases monotonically  with the baryon number density \cite{ser}. This is in 
contrast  to several
variational many-body (VMB) calculations of the equation of state (EOS) of 
nuclear matter \cite{wir}, which predict the symmetry energy to
saturate and then  to decrease at high densities.
The aim of this paper is
to study how sensitive the formation of a mixed quark-nucleon phase
in neutron stars is to the high density behavior
of the nuclear symmetry energy.

It was suggested in Ref. \cite{glen} that the isospin properties of the
RMF model are responsible for the existence of  the mixed
quark-nucleon phase. We show here that it is the behavior of nucleon
matter isobars that allows the existence of the mixed phase,
irrespective of the particular form of the nuclear symmetry energy.

One should stress that different nuclear matter
models, which fit the saturation point, display
a rather diverse high density behavior of the nuclear symmetry energy.
As mentioned already, variational calculations with phenomenological
nucleon-nucleon potentials predict a density dependence that is incompatible
with that of the RMF models. This discrepancy leads to  serious uncertainty
about some astrophysically relevant properties of the neutron star matter. Here we 
study  the consequences of this uncertainty for the nucleon-quark
phase transition in neutron star cores.

\section{Quark matter in neutron stars}
Glendenning's construction describes a global division of the
baryon number
between  two phases. It is insensitive to the geometrical form  of the volume
enclosing, respectively, nucleons and quarks. To account for the shape and  
size of droplets of each phase, one should include the 
surface tension at the interface between the nucleon matter and the quark matter,
the Coulomb interaction, and the Debye screening.
Inclusion of these effects results in some
corrections to the equation of state, which do not alter considerably the 
results 
of this simple approach. We discuss  
finite-size effects in Sec. VI
where both the Coulomb interaction and the surface tension are
included in the calculations.

 The equilibrium conditions, in the case where the geometry of droplets is 
 neglected, are those for bulk systems. 
The neutron star matter is assumed to be  $\beta$ stable and charge
neutral. Thermal effects are not expected to play any important
role in neutron star cores. We neglect them and put the
temperature $T=0$. 

The equilibrium conditions for the quark matter droplet to
coexist with the nucleon medium are that pressure and chemical
potentials in both phases coincide. We choose pressure as an
independent variable. The coexistence requires that

\begin{equation}
\mu_N^n(p)=\mu_N^q(p)  
\end{equation}
 and 
\begin{equation}
\mu_P^n(p)=\mu_P^q(p),  
\end{equation}
 where $\mu_N^n$ and $\mu_N^q$ are the neutron
chemical potentials in the nucleon  and the quark phase,
respectively. Similarly,
$\mu_P^n$ and $\mu_P^q$ are the proton chemical
potentials in respective phases.
To solve the above coexistence conditions we 
construct isobars for both phases of baryon matter.

 The $\beta$-equilibrium
condition reads

\begin{equation} 
\mu_N^i-\mu_P^i=\mu_e,~~~~~~~i=n,q,  
\end{equation}
 where $\mu_e$ is the electron chemical potential, and $n$
and $q$ refer to the nucleon and the quark phases, respectively. It turns
out that muons can  safely be neglected as the electron chemical
potential remains low. We assume  that the electron
distribution is uniform. In this case,
neglecting the electron rest mass, 

\begin{equation} 
\mu_e=(3\pi^2n_e)^{1/3}, 
\end{equation}
 where $n_e$ is the electron density. 

Solutions of Eqs. (1) - (3) provide  the densities of both
baryon phases $n_P$ and $n_q$, where $n_P$ is the proton
density and $n_q$ is the electric charge density of quarks in units of $e$. 
The global charge neutrality condition requires that

\begin{equation} 
Vn_e=V_nn_P+V_qn_q, 
\end{equation}
where $V_n$ is the volume occupied by nucleons and $V_q$ is
the volume of quarks. Since $V_n+V_q=V$, the total
available volume, we can define a quantity $\alpha=V_n/V$, which
is the fraction of space containing nucleons. The quarks occupy
a complementary fraction, $1-\alpha$, of the volume. From Eq. (5)
we obtain $\alpha$ in the form 

\begin{equation} 
\alpha = {n_e-n_q \over n_P-n_q}. 
\end{equation}

At a sufficiently low pressure we expect $\alpha=1$, as free quarks are absent
in the neutron star matter. The first quark
droplets form at some pressure $p_i$, which we refer to as the
lower critical pressure. It corresponds to $\alpha$
starting to deviate from unity 
for the first time. With increasing pressure, more space is
filled with the quark matter and $\alpha<1$. Nucleons disappear
at the upper critical pressure $p_f$, with $\alpha(p_f)=0$. For
pressure in the range  $p_i<p<p_f$,
nucleon matter coexists with quark matter.  
 
The lower and upper critical pressure values $p_i$ and $p_f$ 
depend on both the nucleon matter EOS and
the model of the quark matter. In the next section we specify the quantities
used in our calculations of properties of the mixed phase.

\section{The nuclear symmetry energy at high densities}

Following Ref. \cite{lag}, we adopt a simple parametrization of the
nuclear matter EOS. As shown in Ref. \cite{lag}, the results of variational
many-body calculations with phenomenological nucleon-nucleon
potentials can be simply parametrized as a function of the
baryon number density and the proton fraction.
The energy per particle  can be expressed as

\begin{equation} 
E(n,x)=T(n,x)+V_0(n)+(1-2x)^2V_2(n), 
\end{equation}
where  $n$ is the baryon number density and $x=n_P/n$ is the
proton fraction.
The kinetic energy contribution is

\begin{equation}
T(n,x)={3 \over 5} {1 \over
2m}(3\pi^2n)^{2/3}[(1-x)^{5/3}+x^{5/3}]. 
\end{equation} 
The functions $V_0(n)$ and $V_2(n)$ represent the
interaction energy contributions. 
The form (7) of the energy per particle is a very 
good approximation of the numerical calculations as
far as the $x$ dependence is concerned \cite{lag}. 

 \begin{figure}
 \begin{center}
 \epsfxsize=8cm
 \epsfbox{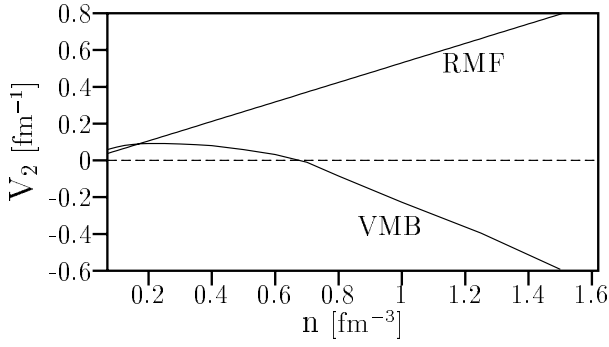}
 \end{center}
 \caption{The interaction energy $V_2(n)$ as a function of the baryon
number density, for the VMB and RMF models.}
 \label{fig1}
 \end{figure}

\vspace*{0.5cm}
From Eq. (7) we obtain the pressure and chemical potentials of
neutrons and protons. 
The pressure is

\begin{equation}
p={2 \over 5} {1 \over 2m}(3\pi^2n)^{2/3}n[(1-x)^{5/3}+x^{5/3}]+n^2V_0'(n)+
(1-2x)^2n^2V_2'(n). 
\end{equation}
The chemical potentials read

\begin{eqnarray}\mu_N^{n}&=&{1 \over 2m}(3\pi^2n)^{2/3}[(1-x)^{5/3}+
  x(1-x)^{2/3}]+V_0(n)+nV_0'(n) \nonumber \\
& & + (1-4x^2)V_2(n)+(1-2x)^2nV_2'(n)+m, 
\end{eqnarray}
\begin{eqnarray}\mu_P^{n}&=&{1 \over 2m}(3\pi^2n)^{2/3}[x^{5/3}+(1-x)x^{2/3}]+
V_0(n)+nV_0'(n) \nonumber \\
& & + (-3+8x-4x^2)V_2(n)+(1-2x)^2nV_2'(n)+m. 
\end{eqnarray}

Our aim here is to study the influence of the nuclear symmetry
energy on the mixed phase properties. The 
energy per particle, Eq. (7), is well suited for this purpose as only
the function $V_2(n)$ enters the expression for the nuclear
symmetry energy, which is

\begin{equation}
E_{sym}(n)= {5 \over 9}T \left(n,{1 \over 2}\right)+V_2(n). 
\end{equation}
To account for the uncertainty in  high density behavior of
$E_{sym}(n)$, we use different parametrizations of $V_2(n)$, keeping 
the function $V_0(n)$ fixed.

As an example of variational many-body calculations, we use
the EOS with the UV14+TNI interactions from 
Wiringa, Fiks, and Fabrocini \cite{wir}. Polynomial fits of this EOS are
given in the Appendix. 
In Fig. \ref{fig1} we show  the function $V_2(n)$ corresponding to the
UV14+TNI interactions. One should note that with this $V_2(n)$ 
the symmetry energy (12) reproduces the empirical value,
$E_{sym}(n_0)=34\pm 4 $ MeV \cite{mye}. At higher densities, $E_{sym}(n)$
saturates and then decreases, reaching negative values for $n>1.0$
fm$^{-3}$. 

The energy per particle of the RMF model can also be cast in the
form (7) \cite{kuts1}. The function $V_2(n)$  in this case is

\begin{equation} 
V_2(n)={1 \over 8} {g_{\rho}^2 \over m_{\rho}^2}n. 
\end{equation}
The coupling parameter $g_{\rho}^2/m_{\rho}^2$ is adjusted
to fit the empirical value of the nuclear symmetry energy
$E_{sym}(n_0)$. The function $V_2(n)$, Eq. (13), grows linearly
with the baryon number density $n$, Fig. \ref{fig1}, and thus
the nuclear symmetry energy in the RMF model increases
monotonically with $n$. As we are concerned here mainly with the role
of the symmetry energy, we model the energy per
particle corresponding to the RMF theory 
using the function $V_2(n)$ in the form (13) and keeping other
contributions in Eq. (7) the same as in the VMB case.

\begin{figure}
 \begin{center}
 \vspace*{0.5cm}
 \epsfxsize=7cm
 \epsfbox{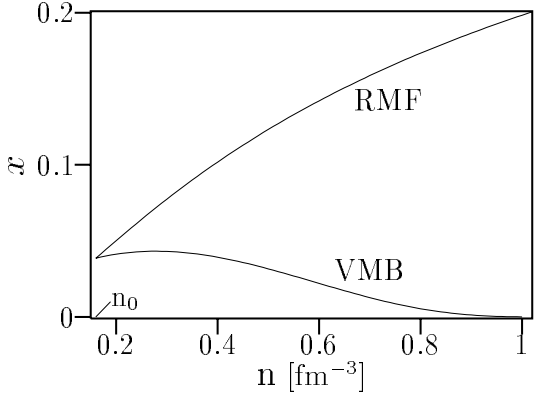}
 \end{center}
 \caption{The proton fraction of the $\beta$-stable neutron star matter 
 corresponding
 to the interaction energy $V_2(n)$ in Fig. 1, for VMB and RMF models.}
 \label{fig2}
 \end{figure}

The $\beta$-equilibrium condition (3) and the formulas (10) and
(11) show that the proton fraction of the neutron star matter is
fully determined by the function $V_2(n)$. In Fig. \ref{fig2} we show the
proton fraction corresponding to both forms of $V_2(n)$.
For both curves, the proton fraction at $n_0$ is
$x\approx 0.05$. This is due to the empirical value
$E_{sym}(n_0)$, which is reproduced by both forms of $V_2(n)$. At
higher densities, the behavior of $x(n)$ is different in the two
cases. The RMF model predicts  that $x(n)$ monotonically increases
with the density, whereas for $V_2(n)$ corresponding to the UV14+TNI
interactions, the proton fraction decreases with $n$, and
eventually protons disappear completely at some density $n_v$,
$x(n_v)=0$. The disappearance of the proton fraction at high densities is 
a general property of nuclear interaction models that give 
$V_2(n)<0$ \cite{kuts2}.

\vspace*{0.5cm}
\section{Quark and nucleon isobars}
The quark matter is described by a simple bag-model equation of state.
We use the same parameters as in Ref. \cite{heis}. The energy density
for three flavors is, neglecting bare masses,

\begin{equation}
\epsilon_q={3 \over
4}\pi^{2/3}(n_u^{4/3}+n_d^{4/3}+n_s^{4/3})+B, 
\end{equation}
where $n_i$, $i=u,d,s$ are the quark number densities.
The quark chemical potentials are

\begin{equation}
\mu_i=\pi^{2/3}n_i^{1/3},~~~~~~~i=u,d,s. 
\end{equation}
The pressure of  quark matter reads

\begin{equation}
 p={1 \over 4\pi^2} (\mu_u^4+\mu_d^4+\mu_s^4)-B.
 \end{equation}
Below we show results corresponding to two values of the bag
constant,
$B=120$ MeV/fm$^3$ and
$B=200$ MeV/fm$^3$.
 We choose these values in order to assess
the sensitivity of the mixed phase properties
to the bag constant, which is treated here as a phenomenological
parameter subject to a substantial uncertainty.  

The $\beta$ equilibrium requires that the chemical
potential of down and strange flavors is the same,

\begin{equation}
\mu_d=\mu_s, 
\end{equation}
and the chemical potential of up quarks satisfies the
condition 

\begin{equation}
\mu_d=\mu_u+\mu_e. 
\end{equation}
Proton and neutron chemical potentials in the quark phase are

\begin{equation}
\mu_P^q=2\mu_u+\mu_d, 
\end{equation}

\begin{equation}
\mu_N^q=\mu_u+2\mu_d. 
\end{equation}

The simplest way to solve the equilibrium conditions (1) and (2) is to
construct isobars for the nucleon and quark matter in the
$\mu_P-\mu_N$ plane. This can be done easily for the nucleon 
matter since proton and neutron
chemical potentials, for a given value of 
pressure, are parametrized by one independent variable. For the
quark phase, we impose the condition (17), which reduces the number of
independent quark chemical potentials, for a given pressure, to
two.  Hence for  
each phase one can draw an isobar, a parametric curve in the $\mu_P-\mu_N$ 
plane corresponding to a fixed pressure.

 \begin{figure}
\vspace*{-0.3cm}
 \hspace*{0.2cm}
 \begin{minipage}{7.8cm}
 \epsfxsize=7cm
 \epsfbox{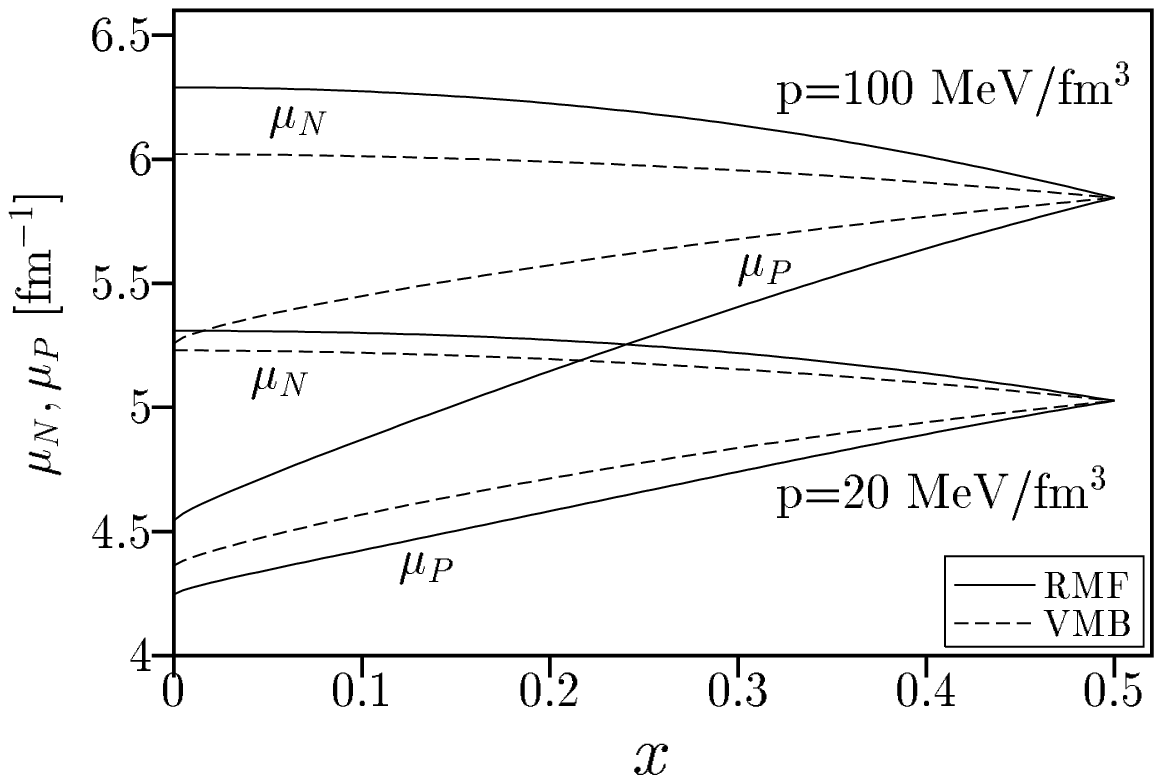}
 \vspace*{0.2cm}
 \caption{Proton and neutron chemical potentials of  nucleon matter
as functions of the 
proton fraction for indicated values of pressure, for VMB and
RMF models.}
 \label{fig3}
 \end{minipage}
\hspace*{0.7cm}
 \begin{minipage}{7.8cm}
 \vspace*{0.61cm}
 \epsfxsize=7cm
 \epsfbox{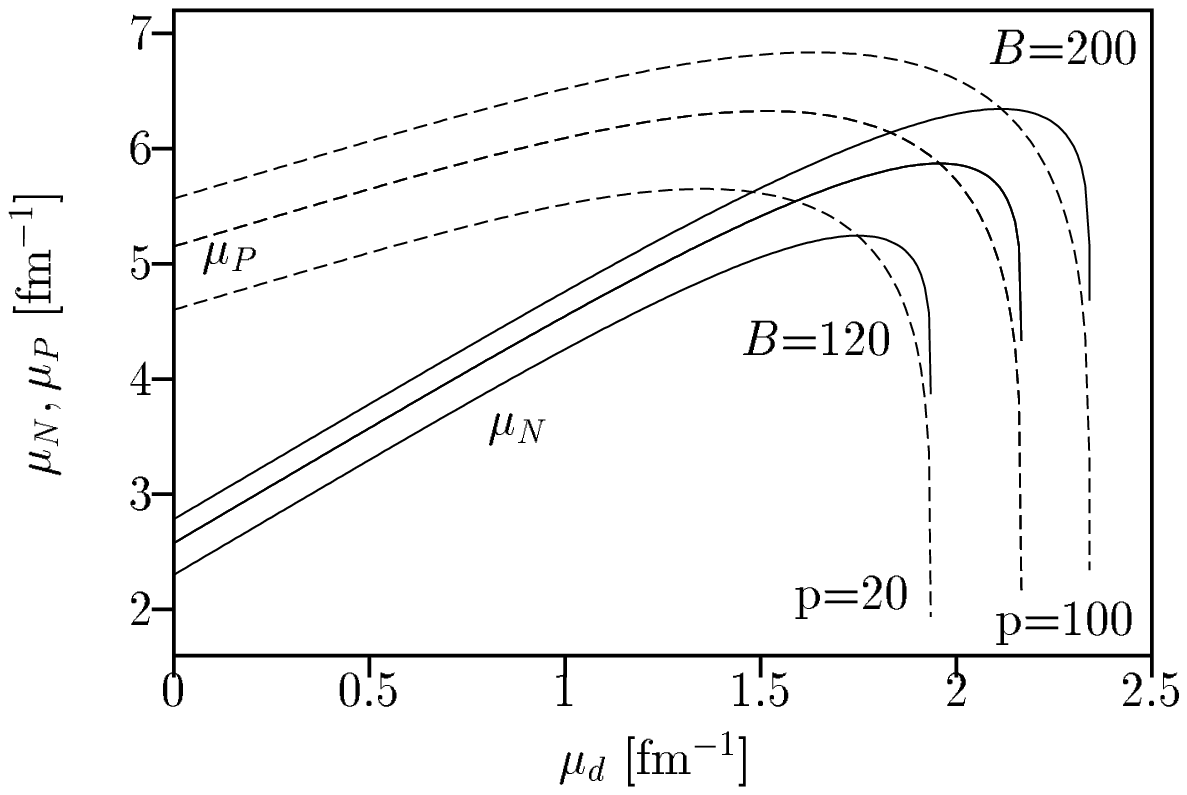}
  \vspace*{0.2cm} 
 \caption{Proton and neutron chemical potentials in quark matter as
functions of the down-quark chemical potential.
Values of pressure and the bag constant  are in MeV/fm$^3$. Solid
and dashed curves correspond, respectively, to neutron and
proton chemical potentials.}
 \label{fig4}
 \end{minipage} 
 \end{figure}

\vspace*{0.3cm}  
For the
nucleon phase we parametrize isobars by the proton fraction $x$. 
As an example, in Fig. \ref{fig3} we show proton and neutron chemical
potentials as functions of $x$ for two values of pressure, 
$p=20$ MeV/fm$^3$ and $p=100$ MeV/fm$^3$. One can see that for 
$p=20$ MeV/fm$^3$ the proton and neutron chemical potentials
corresponding to VMB and RMF models do not differ much from one another. For 
$p=100$ MeV/fm$^3$, the VMB and RMF curves deviate significantly. In 
particular, 
the difference $\mu_N(p,x=0)-\mu_P(p,x=0)$ decreases with pressure for the 
VMB model, whereas it increases for the RMF one. 

In the quark phase, using the condition (17), one finds an isobaric relation
between up and down quark chemical potentials in the form

\begin{equation}
\mu_u^4=4\pi^2(p+B)-2\mu_d^4. 
\end{equation}
We choose $\mu_d$ as an independent variable. In Fig.
\ref{fig4} the proton and neutron chemical potentials in the quark matter are
shown for the same values of pressure as above, $p=20$ MeV/fm$^3$ and 
$p=100$ MeV/fm$^3$, for both values of the bag constant, $B=120$ MeV/fm$^3$ 
and $B=200$ MeV/fm$^3$. One can see that the curve corresponding to $p=20$
 MeV/fm$^3$ 
and $B=200$ MeV/fm$^3$ coincides with that for $p=100$ MeV/fm$^3$ and 
$B=120$ MeV/fm$^3$.

 \begin{figure}
\vspace*{0.5cm}
 \hspace*{0.2cm}
 \begin{minipage}{7.8cm}
 \epsfxsize=7cm
 \epsfbox{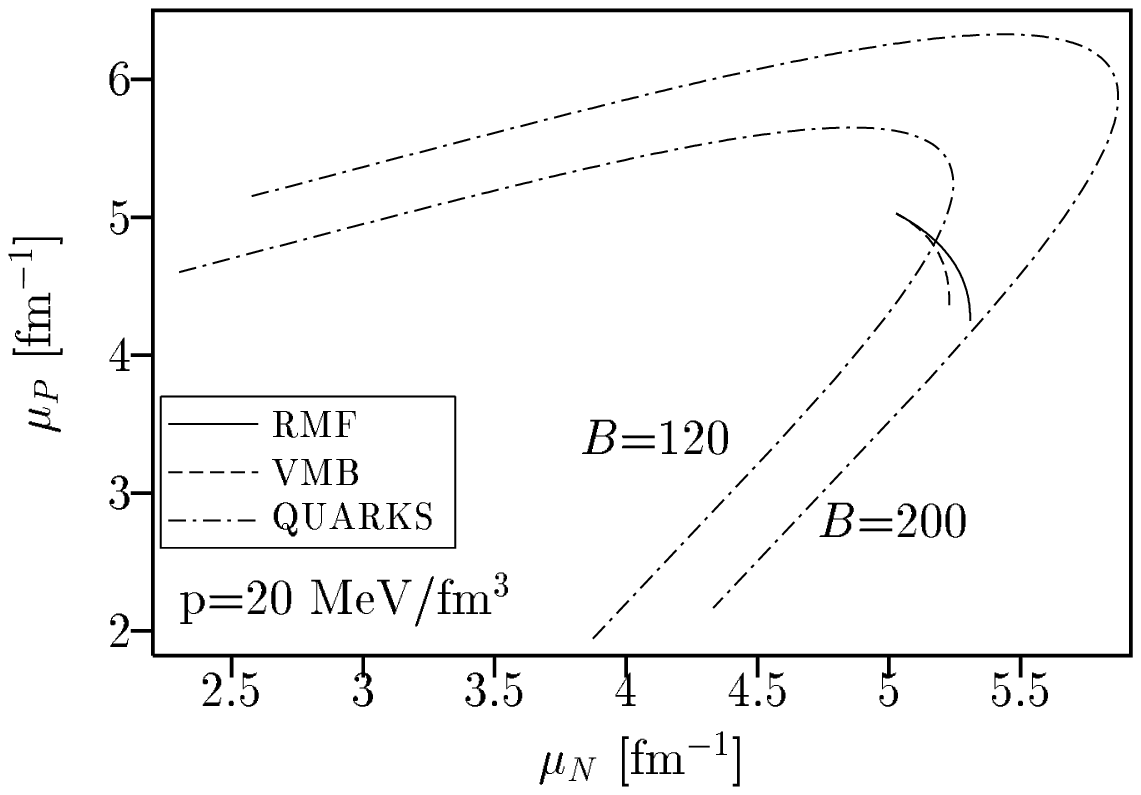}
 \vspace*{0.2cm}
 \caption{VMB and RMF nucleon isobars, and quark isobars for two
values  of the bag constant, for pressure $p=20$ MeV/fm$^3$.}
 \label{fig5}
 \end{minipage}
\hspace*{0.7cm}
 \begin{minipage}{7.8cm}
\vspace*{-0.4cm}
 \epsfxsize=7cm
 \epsfbox{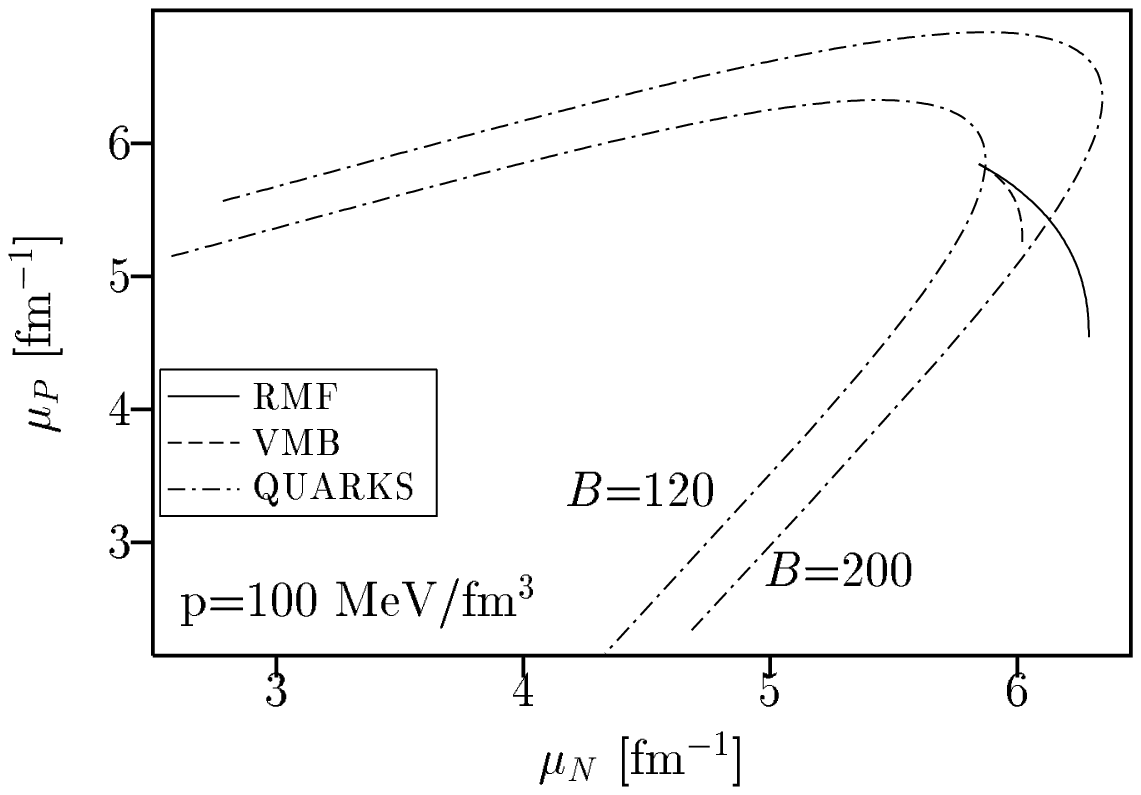}
  \vspace*{0.2cm} 
 \caption{The same as in Fig. 5 for pressure $p=100$ MeV/fm$^3$.}
 \label{fig6}
 \end{minipage} 
 \end{figure}

Isobars for the two phases are shown in Figs. \ref{fig5} and \ref{fig6},
 where $\mu_P$ is 
plotted 
against $\mu_N$  for $p=20$ MeV/fm$^3$ and $p=100$ MeV/fm$^3$, respectively. 
The coexistence conditions (1) and (2) are fulfilled at 
the crossing point of the nucleon and quark isobars on
the $\mu_P-\mu_N$ plot. Also, the $\beta$-equilibrium
condition (3) is satisfied at this point with the electron
chemical potential $\mu_e=\mu_P-\mu_N$. This formula allows us
to calculate the  density of the homogeneous electron background
$n_e$ from Eq. (4).

The quark matter isobars are calculated assuming the strange
quark mass to vanish, $m_s=0$. This approximation is justified
for pressure values relevant to our analysis, as in this range
isobars corresponding to $m_s=0$ and to the empirical value of 
the strange quark mass, $m_s =150$ MeV, are similar. 
Also, the lowest values of the strange quark chemical potential
obtained  from the coexistence conditions (1) and (2), 
$\mu_s \sim 400$ MeV, exceed  the empirical mass $m_s=150$
MeV considerably. Corrections due to the empirical value of the strange quark
mass would result in small changes of the critical parameters
corresponding to the formation of the first quark droplets.

\section{Results and implications for neutron stars}
In order to find a critical pressure indicating the onset of the phase
transition to quark matter, we analyze how the crossing of nucleon
and quark isobars changes with pressure. 
For very low values of pressure, $p \le 1$ MeV/fm$^3$, the nucleon
isobars for both VMB and RMF models do not cross the quark isobars, for both
values of the bag constant. With increasing pressure, the lower
end point of the nucleon isobar, corresponding to $x=0$, merges
with the quark isobar at some  pressure value $p_0$.
At this pressure  pure neutron matter can coexist with  quark
matter. This, however, is not the situation encountered in
neutron stars, where at this value of pressure the neutron star
matter contains a small proton fraction of about $5\%$, as shown
in Fig. \ref{fig2}. For higher pressure, $p>p_0$, the proton fraction at
the crossing point of the nucleon isobar with the quark isobar
increases. The formation of the first quark droplets
in the nucleon medium starts at such a pressure $p_i$
that the proton fraction of the nucleon matter at the
crossing of the isobars 
coincides with that of the $\beta$-stable neutron star matter at
this pressure. The pressure $p_i$ is referred to as the lower
critical pressure.

For $B=120$ MeV/fm$^3$, the lower critical pressure for the VMB
and RMF isobars is, respectively, 
$p_i=2$ MeV/fm$^3$ and $p_i=3$ MeV/fm$^3$.
The upper critical pressure $p_f$ at which the last nucleon
droplets immersed in  quark matter finally dissolve, is the same for 
both VMB and RMF isobars, $p_f=115$ MeV/fm$^3$ (see Fig. \ref{fig6}).

In the case of $B=200$ MeV/fm$^3$, the lower critical pressure for VMB and
RMF models is, respectively, $p_i=215$ MeV/fm$^3$ and $p_i=35$
MeV/fm$^3$. The value of the upper critical pressure is $p_f=290$
MeV/fm$^3$.

\begin{figure}
 \vspace*{0.1cm}
 \begin{center}
 \epsfxsize=7cm
 \epsfbox{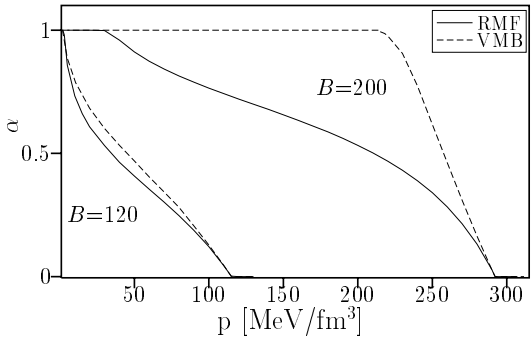}
 \end{center}
  \caption{The fraction of the mixed phase volume filled with nucleons
as a function of pressure for VMB and RMF models and for both values of $B$.}
 \label{fig7}
 \end{figure}

The  fraction of  volume filled with nucleons, $\alpha$, is shown in Fig. 
\ref{fig7}.
For the lower value of the bag constant, $B=120$ MeV/fm$^3$, the
curves for VMB and RMF models are very similar. This is because
the lower critical
pressure $p_i$ is almost identical in the two cases. The situation
is quite different for $B=200$
MeV/fm$^3$. In this case the phase transition for the VMB model
starts at much higher pressure than for the RMF model. This
reflects the fact that the VMB isobar 
becomes much shorter at high pressure than the RMF one. One
can easily see this difference in Fig. \ref{fig6} where we show $p=100$
MeV/fm$^3$ isobars. 

\vspace*{0.2cm}
The results shown in Fig. \ref{fig7} prove that the
properties of the nucleon-quark phase transition are very sensitive
to the nuclear symmetry energy, for higher values of the bag
constant. This is because, generally, the phase transition occurs
at a higher pressure for higher values of the bag constant, and the
VMB and RMF isobars differ much more at high values of pressure.
For low values of $B$, the phase transition starts at a low enough
pressure for the nuclear symmetry energy not to affect the
isobars significantly.

 \begin{figure}
 \vspace*{0.1cm}
 \begin{center}
 \epsfxsize=7cm
 \epsfbox{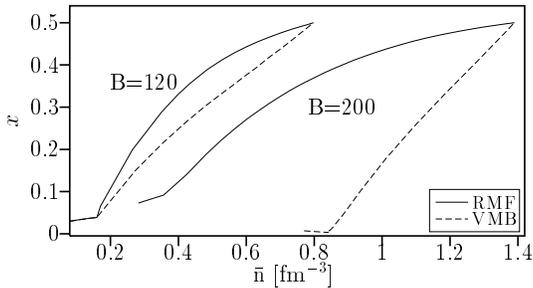}
 \end{center}
 \caption{The proton fraction of nucleon matter coexisting with 
quark matter as a function of the mean baryon number density.}
 \label{fig8}
 \end{figure}

\vspace*{0.2cm}
In Fig. \ref{fig8} we show the proton fraction of  nucleon matter
coexisting at a given pressure with quark matter, as a
function of the mean 
baryon number density $\bar n$,

\begin{equation} 
{\bar n}=\alpha n + (1-\alpha)n^Q, 
\end{equation}
where $n$ and $n^Q=(n_u+n_d+n_s)/3$ are the baryon number
densities of, respectively, nucleon and quark matter. 
The density corresponding to the lower critical pressure $p_i$ at
which the phase transition starts is ${\bar n}_i$.
The phase
transition is completed at the density ${\bar n}_f$ corresponding to
the upper critical pressure $p_f$. For $B=120$ MeV/fm$^3$, ${\bar n}_i=0.17$
fm$^{-3}$ is approximately the same for VMB and RMF models.
One can notice that nucleon matter becomes more proton rich
with increasing pressure irrespective of the nuclear symmetry
energy. At the upper critical pressure $p_f$ disappearing nucleon
droplets for both nuclear models are  composed of symmetric nuclear matter.  
The corresponding density is ${\bar n}_f=0.8$ fm$^{-3}$.
For $B=200$ MeV/fm$^3$ the quark droplets start to form at a density 
${\bar n}_i=0.84$ fm$^{-3}$ and ${\bar n}_i=0.35$ fm$^{-3}$, respectively, 
for the VMB and RMF models. Nucleons disappear at ${\bar n}_f=1.39$ fm$^{-3}$.

 \begin{figure}
 \vspace*{0.1cm}
 \begin{center}
 \epsfxsize=7cm
 \epsfbox{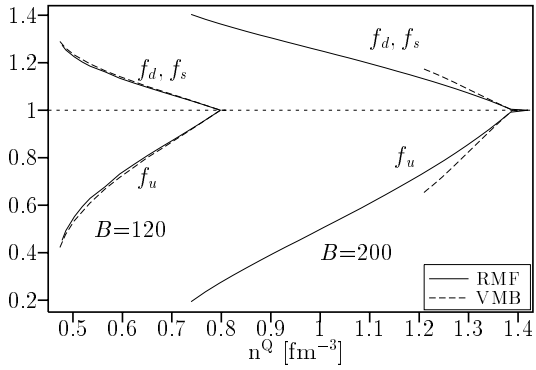}
 \end{center}
 \caption{Concentrations of quark flavors in  quark matter
coexisting with  nucleon matter as functions of the quark
baryon density.}
 \label{fig9}
 \end{figure}

\vspace*{0.2cm}
Properties of quark matter in the mixed phase are displayed in Fig. \ref{fig9},
 where 
the flavor composition is shown for both values of the bag
constant. The flavor  
concentrations are $f_i=n_i/n^Q$, $i=u,d,s$. A strong dependence
of the mixed phase properties on the symmetry
energy  for $B=200$ MeV/fm$^3$ is best visible in 
this figure. Quark matter, forming  the first droplets at the
lower critical pressure
$p_i$, is composed mostly of negatively charged quarks. With increasing pressure
the flavor composition becomes more symmetric, and, at the upper
critical pressure $p_f$, the 
concentrations of all flavors become equal, $f_u=f_d=f_s$. Quark matter at
$p \ge p_f$ is charge neutral. 

\begin{figure}
 \begin{center}
 \epsfxsize=7cm
 \epsfbox{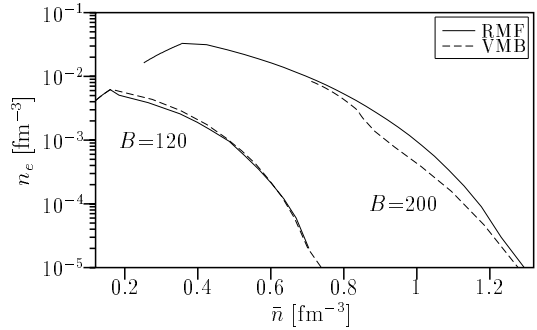}
 \end{center}
 \caption{The  electron density of the uniform electron background as
a function of the mean baryon number density.}
 \label{fig10}
 \end{figure}

The electron density is shown in Fig. \ref{fig10}. At the lower critical 
pressure $p_i$, 
$n_e=n_P$, as only electrons compensate the positive charge carried by protons.
With increasing pressure, negatively charged quarks become more abundant and the
electron density decreases. Electrons disappear at the pressure $p_f$ when the
whole available volume is filled with electrically neutral  quark matter.

\begin{figure}
\vspace*{0.7cm}
 \hspace*{0.2cm}
 \begin{minipage}{7.8cm}
 \epsfxsize=7cm
 \epsfbox{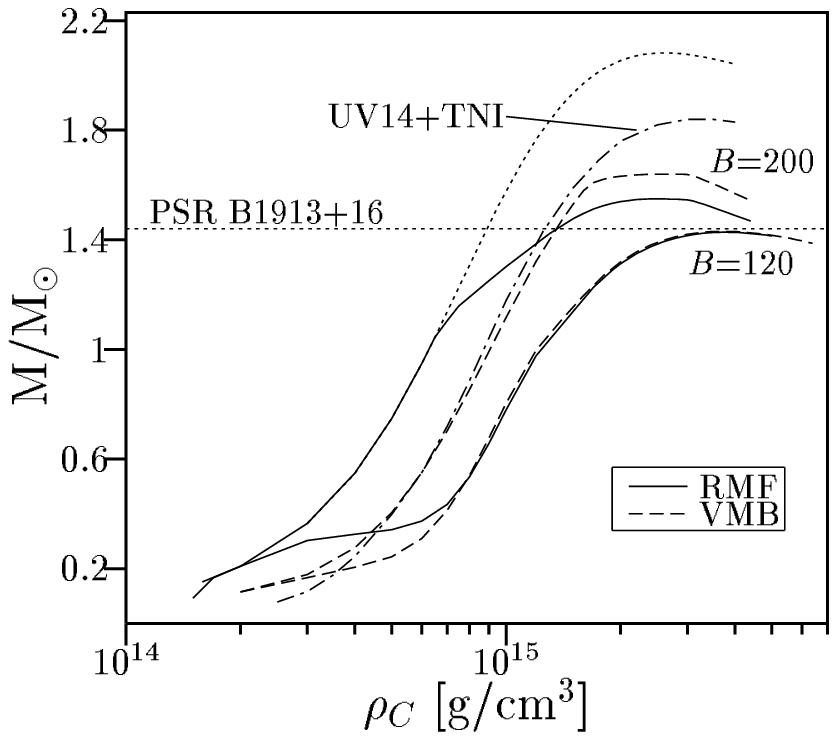}
 \vspace*{0.2cm}
 \caption{Neutron star masses as functions of the central density. The
dotted line corresponds to pure nucleon matter in the RMF model.
The dash-dotted line is for pure nucleon VMB equation of
state. Solid and dashed curves are for equations of state
involving the mixed quark-nucleon phase.
The horizontal line shows the empirical lower limit to the
maximum neutron star mass.}
 \label{fig11}
 \end{minipage}
\hspace*{0.7cm}
 \begin{minipage}{7.8cm}
\vspace*{-0.3cm}
 \epsfxsize=7cm
 \epsfbox{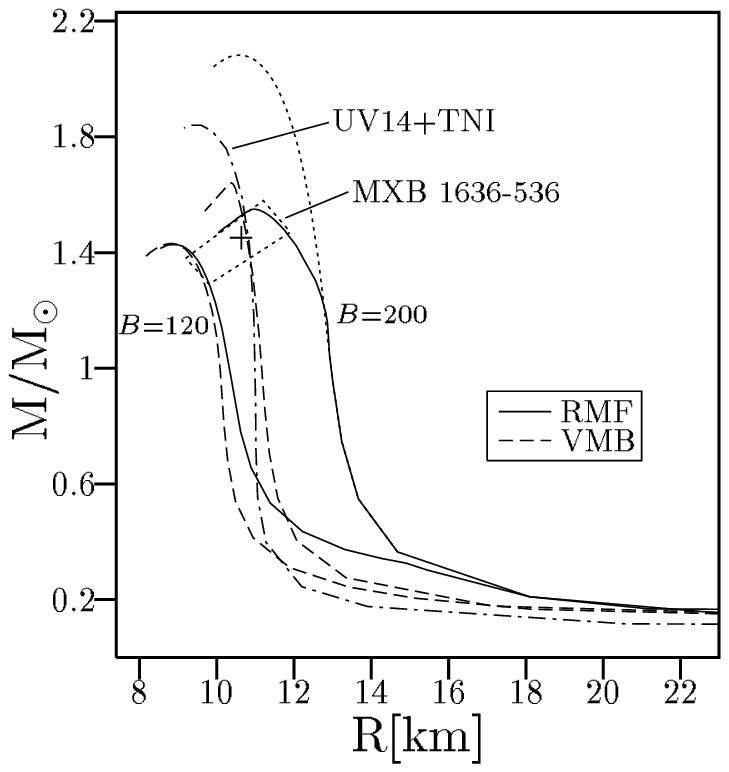}
  \vspace*{0.2cm} 
 \caption{Neutron star masses as functions of the radius for the same
equations of state as in Fig. 11. The cross and the error box
represent empirical constraints for the x-ray source MXB 1636-536.}
 \label{fig12}
 \end{minipage} 
 \end{figure}

\vspace*{0.4cm}
To investigate the consequences  of the existence of a mixed 
quark-nucleon phase for neutron stars we construct the equation
of state $p=p({\bar \rho})$,  
where the mass density of the mixed phase is 

\begin{equation} 
{\bar \rho}={1 \over c^2}[\alpha \epsilon_{nuc}+(1-\alpha)\epsilon_q]. 
\end{equation}
In Fig. \ref{fig11}  masses of neutron stars are shown as functions of 
the central
density for  pure nucleon matter and for the 
mixed quark-nucleon phase.
The phase transition to quark matter makes the equation of state softer.
The maximum mass of neutron stars containing the mixed phase decreases 
as compared with the pure nucleon case. The effect is stronger for low values 
of the bag constant, as the mixed phase comprises more mass of the star
than for higher $B$.
Results shown in Fig. \ref{fig11} exclude, in fact, low values of the bag 
constant, as for
$B \le 120$ MeV/fm$^3$ the maximum mass is below the observational limit.  For 
$B=200$ MeV/fm$^3$ the maximum mass safely exceeds this limit. In this case, 
the influence of the nuclear symmetry energy is clearly visible. The maximum 
mass corresponding to the RMF model is well below that for the VMB model. 
Generally,
reduction of the maximum mass due to the presence of the mixed phase is 
much more
significant for the RMF model than for the VMB one. This reflects the fact that
for $B=200$ MeV/fm$^3$ the phase transition to quark matter starts at much lower
pressure in the RMF model than in the 

\newpage
\noindent VMB case. Correspondingly,
the mixed phase core of the maximum mass neutron star is much
smaller for the VMB model than for the RMF one.

We also show in Fig. \ref{fig12} neutron star masses as functions of the
radius for the same EOS's as in Fig. \ref{fig11}. The error box indicates
the constraints on the mass-radius relation for the x-ray
burst source MXB 1636-536. In this case, equations of state
involving a mixed phase with the bag constant $B=120$ MeV/fm$^3$ are
marginally compatible with the data.

It is interesting to note that the 
structure of a neutron star of the
canonical mass $M=1.44M_{\odot}$ is even more sensitive to the form of 
the nuclear symmetry
energy. For $B=200$ MeV/fm$^3$, with the RMF symmetry energy, such
a star, whose radius is $\sim 12$ km, possesses a mixed phase core
of  $\sim 7$ km radius. In the VMB  model, the star is composed entirely
of nucleons. The central pressure of the VMB star, $p_c=120$
MeV/fm$^3$, is below the lower critical value
$p_i=215$ MeV/fm$^3$ at which the phase transition to the  
mixed quark-nucleon matter begins.

\section{Coulomb and surface effects}
Up to now we did not consider the space structure of the mixed
phase. When the Coulomb interaction and the surface tension are
included in the calculations, one can find a variety of
geometric structures formed by regions filled with nucleons and
quarks that have opposite electric charge density \cite{rav}.
The form of these structures
changes with the fraction of space filled with nucleons,
$\alpha $. It should be stressed, however, that the mixed
phase (called also, following Ref. \cite{heis}, the droplet phase)
is the ground state of the neutron star matter only if the
surface tension at the  
interface between nucleon and quark matter remains sufficiently small 
\cite{heis}.
In the opposite case, the mixed phase will not be favored energetically and a
phase transition will lead to the coexistence of two bulk electrically neutral
phases. We investigate here how the high density behavior of the nuclear symmetry
energy influences the Coulomb and surface effects
in the droplet phase. 

We perform calculations using the Wigner-Seitz approximation. In this
approach we assume that characteristic sizes of structures in the droplet phase
are less than the Debye screening length, which is about $10$ fm for the nucleon
phase and $5$ fm for quarks \cite{heis}. In this case electrons are
essentially uniformly distributed
over the whole system and so are other particles within a given
phase. When the spatial scale of structures
is larger than the screening length the charge density becomes
nonuniform and the mixed phase resembles two coexisting neutral
phases. 

The geometric forms of structures considered here are
droplets or bubbles, rods, plates, and also intermediate forms, each of them
characterized by a continuous dimensionality parameter $d$
ranging from $d=3$ for droplets to 
$d=1$ for plates. For all these geometries the minimized sum of Coulomb and 
surface energy densities reads

\begin{equation}
\epsilon _C + \epsilon _{\sigma } = 6\pi \chi \left( {\sigma ^{2}d^{2} 
[n_{P}(\alpha )-n_{q}(\alpha )]^{2}e^{2}f_{d}(\chi) \over
16\pi ^{2}} \right) ^{1/3}.  
\end{equation}
The radius of the rare phase bubble immersed in the dominant
phase is

\begin{equation}
r = \left(4\pi {[n_{P}(\alpha )-n_{q}(\alpha )]^{2}e^{2}f_{d}(\chi) \over
\sigma d}\right)^{-1/3}.
\end{equation}
Here $n_{P}(\alpha)$ and $n_{q}(\alpha)$ are charge densities of nucleons 
and quarks, in 
units of $e$, corresponding to a given proportion of phases $\alpha$.
The quantity $r$ expresses the characteristic sizes of geometric forms and 
particularly
for droplets and rods is their radius, and in the case of plates is their half 
thickness. In the above equations $\chi $ is the fraction of volume occupied
by the rare phase. It is equal to $1-\alpha $ when nucleons are the dominant
phase $(\alpha \geq 1/2)$ and quarks form different structures, and it is simply
$\alpha $ in the opposite situation $(\alpha \leq 1/2)$. It also defines
the half  distance between structures, $R$, which is the radius
of the Wigner-Seitz
cell. The fraction $\chi$ can be expressed in terms of the two
scales $r$ and $R$ as

\begin{equation}
\chi = (r/R)^{d}.
\end{equation}
The function $f_{d}(\chi )$ is

\begin{equation}
 f_{d}(\chi) = {1 \over d+2} \left( {1 \over d-2} \left( 2-d\chi ^{1-2/d}
  \right) + \chi \right). 
\end{equation}
It has a correct logarithmic limit for $d = 2$.

Having the values of $n_{P}$ and $n_{q}$  from 
Eqs. (1)-(3), and (5) we are able to calculate finite-size effects in
the quark-nucleon mixed phase. The values of both $\epsilon _{C} + 
\epsilon _{\sigma }$
and $r$ are obtained by  minimization of the energy, Eq. (24), with respect
to $d$ for a given proportion of phases $\alpha $ and for fixed $\sigma $. 
Because
the exact value of the surface tension is unknown we keep it
as in the work of 
Heiselberg {\em et al.}~\cite{heis} as a parameter which for simplicity is density 
independent.

For the mixed phase to be favorable, its energy density must
be less than the energy densities of all other phases of baryon
matter. When $\sigma  = 0 $ and Coulomb and surface  
effects are absent the mixed phase is energetically favored, as
we showed in Sec. IV by performing a proper
construction of the phase transition. Nevertheless, when $\sigma \neq 0$
the situation is different and for some high values
of the surface tension the mixed phase becomes energetically unfavorable.
Therefore to see what is the ground state of neutron star matter one should
compare the energy density of the droplet phase with the 
energy densities of nucleon matter, quark matter, and  coexisting electrically 
neutral phases of nucleon and quark matter.

 \begin{figure}
\vspace*{0.4cm}
 \hspace*{0.2cm}
 \begin{minipage}{7.8cm}
 \begin{center}
 \epsfxsize=6cm
 \epsfbox{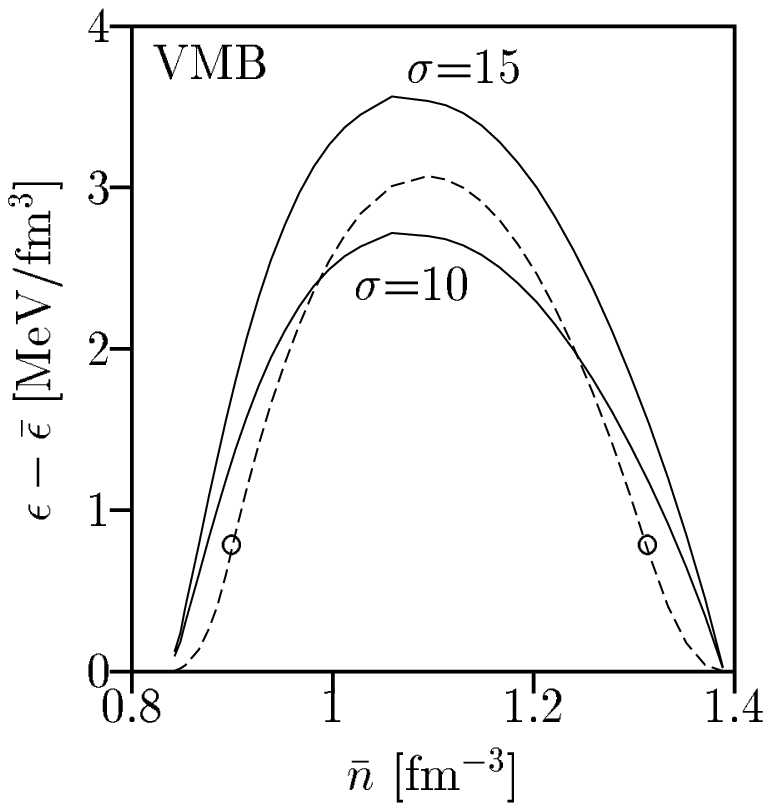}
 \end{center}
 \vspace*{0.2cm}
 \caption{The energy density of the mixed phase in the VMB
model, for indicated values of surface tension $\sigma$ (in MeV/fm$^2$),
relative to its value for 
$\sigma=0$ (solid curves). The dashed curve is the energy gain
in the mixed phase with $\sigma=0$, with respect to electrically
neutral nucleon matter, quark matter and coexisting phases of
nucleon and quark matter. Open circles correspond to bulk
neutral nucleon matter and quark matter that can coexist in the
first order phase transition.}
 \label{fig13}
 \end{minipage}
\hspace*{0.7cm}
 \begin{minipage}{7.8cm}
\vspace*{-3.1cm}
\begin{center}
 \epsfxsize=6cm
 \epsfbox{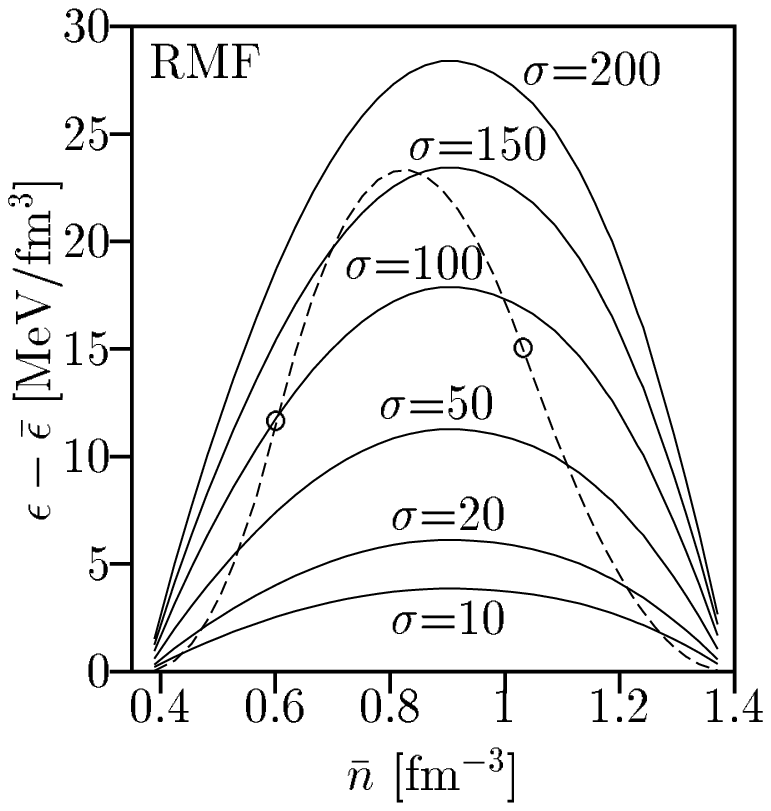}
 \end{center}
  \vspace*{0.0cm} 
 \caption{The same as in Fig. 13 for the RMF model.}
 \label{fig14}
 \end{minipage} 
 \end{figure}

\vspace*{0.4cm}
To account
for the energy density of coexisting phases of electrically
neutral nucleon matter and quark matter we perform a double-tangent 
construction
of the phase transition from nucleon to quark matter.
The energy density gain in the mixed phase with respect to
nucleon matter, quark matter, and coexisting neutral phases of
nucleon and quark matter, $\epsilon - \bar{\epsilon}$, where
$\bar{\epsilon}$ is the
energy density of the mixed phase for $\sigma = 0$, is shown
as a dashed curve in Fig. \ref{fig13} and Fig. \ref{fig14}  for VMB and 
RMF models,  respectively. 

We show here results only
for the bag constant $B = 200$ MeV/fm$^{3}$ as possibly realized in nature. 
As one sees, the difference in energy densities is strongly 
dependent on the nuclear symmetry energy. For the VMB model it is only a few
MeV/fm$^{3}$ whereas it is almost $25$ MeV/fm$^{3}$ for the RMF model. 
This implies
that the allowed range of finite-size effects for the droplet phase to be
energetically favorable will be smaller in the former case and
larger in the latter.
In Figs. \ref{fig13} and \ref{fig14}
the solid curves represent the contributions of Coulomb and surface effects to the
energy density of the droplet phase,  for the indicated values of the surface 
tension. 
If the solid curve for a given $\sigma $ lies below the dashed one
in some range of the baryon number density $\bar{n}$, then
the droplet phase will be the ground state of neutron star matter. If not,
the droplet phase will not be preferred energetically and transition between
two electrically neutral phases will occur. As is shown, our results yield
$\sigma \leq 10$ MeV/fm$^2$ for the VMB model and $\sigma \leq 150$
 MeV/fm$^{2}$ for the RMF one. We can thus conclude  that the appearance 
of the mixed phase, although
dependent on the exact value of the surface tension, is also very sensitive
to the form of the nuclear symmetry energy.

\begin{figure}
 \hspace*{0.2cm}
 \begin{minipage}{7.8cm}
 \begin{center}
 \epsfxsize=5cm
 \epsfbox{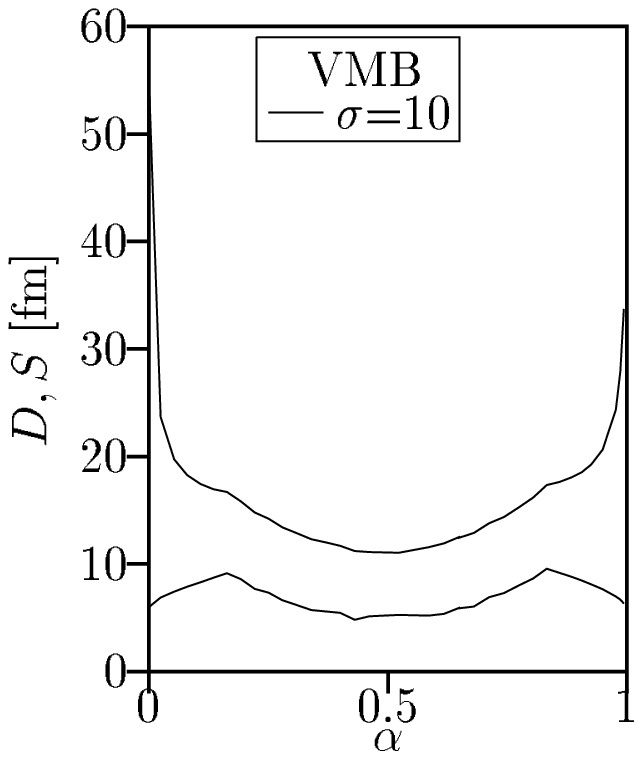}
 \end{center}
 \vspace*{0.2cm}
 \caption{Diameter (lower curve) and separation (upper curve) of structures 
in the mixed phase for the VMB model. The 
surface tension is $\sigma=10$ MeV/fm$^2$.}
 \label{fig15}
 \end{minipage}
\hspace*{0.7cm}
 \begin{minipage}{7.8cm}
\vspace*{-0.4cm}
\begin{center}
 \epsfxsize=5cm
 \epsfbox{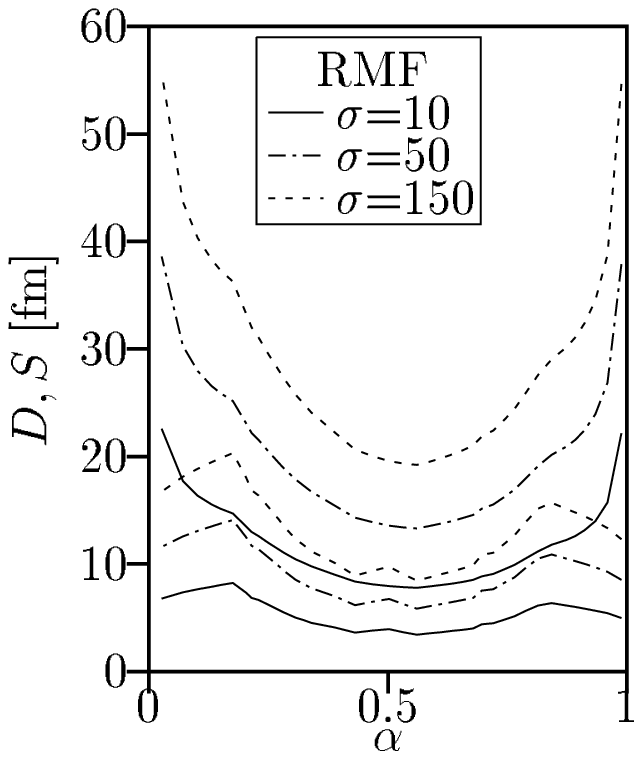}
 \end{center}
  \vspace*{0.2cm} 
 \caption{The same as in Fig. 15 for the RMF model and for
$\sigma=10, 50$, and $150$ MeV/fm$^2$. }
 \label{fig16}
 \end{minipage} 
 \end{figure}

The above calculations were made neglecting the Debye 
screening, which is justified for small size of the structures.
In Fig. \ref{fig15} and Fig. \ref{fig16} we show typical diameters of bubbles
and their separations for the VMB and RMF model, respectively. 
As one sees in Figs. \ref{fig15} and \ref{fig16}, for higher values 
of the surface tension the characteristic sizes of structures
 
\vspace*{0.4cm}
\noindent are larger than screening lengths in nucleon and quark matter so
that the assumption of uniform charge distribution may not be valid.

\section{Discussion}
We have studied the formation of a mixed quark-nucleon phase in
neutron stars, for different models of the nuclear symmetry
energy. We conclude that  the behavior of nucleon matter isobars in the
$\mu_P-\mu_N$ plane, as, e.g., in Figs. \ref{fig5}  and \ref{fig6}, is
fully responsible for the existence of the mixed phase. The isobars, for
any pressure, have
a common end point, corresponding to $x=1/2$, which is thus
independent of the nuclear symmetry energy. For $x<1/2$ the
isobars corresponding to various forms of the nuclear symmetry
energy differ considerably. The most different is the location
of the $x=0$ end point of isobars (see Figs. \ref{fig5} and \ref{fig6}).

The lower critical pressure $p_i$ for the first quark droplets
to form is determined by the location of the $x=0$ end point of a
nucleon isobar, which strongly depends on the form of the nuclear
symmetry energy. The upper critical pressure $p_f$ for the last
nucleons to dissolve into quarks is the same for all isobars
that differ only in the form of the function $V_2(n)$
determining the nuclear symmetry energy. This explains why 
nucleon matter coexisting with  quark matter becomes proton
rich with increasing pressure, and
disappearing nucleon matter at the critical pressure $p_f$ is
symmetric. In Ref. \cite{glen} the increase of the proton fraction
of nucleon matter coexisting in the mixed phase was attributed
to the particular form of the nuclear symmetry energy in the RMF
model.

Properties of neutron stars depend on both lower and upper
critical pressure values, $p_i$ and $p_f$. Since the lower critical
pressure $p_i$ strongly depends on the nuclear symmetry energy,
properties of neutron stars also depend on it. In particular, the
size of the mixed phase core, which determines the maximum mass
of a neutron star, is rather sensitive to the nuclear symmetry
energy. Conclusions concerning the presence of the mixed
quark-nucleon phase in
neutron stars are thus subject to some uncertainty that reflects 
incompatible high density predictions of the nuclear symmetry energy.

When the Coulomb interaction and the surface tension are
included in the calculations, one can determine the 
surface tension values for which the mixed quark-nucleon phase
is the ground state of neutron star matter. As shown in Fig. \ref{fig13}
and Fig. \ref{fig14} the allowed range of $\sigma$ is very narrow for
the VMB model ($\sigma < 10$ MeV/fm$^2$), whereas it is much wider
for the RMF model ($\sigma < 150$ MeV/fm$^2$).

\section*{acknowledgment}

This research was partially supported by the Polish State Committee 
for Scientific Research (KBN), under Grant No. 2 P03B 112 17.

\newpage
\section*{appendix}

We fit the interaction energies $V_0(n)$ and $V_2(n)$ from
Ref. \cite{wir} corresponding to the UV14+TNI interactions with the 
following polynomials:

\begin{equation}
\eqnum{A1}
 V_{0}(n)=-0.0827n^{4}-0.3111n^{3}+2.2624n^{2}-1.181n-0.0571 
 \end{equation}
and

\begin{equation}
\eqnum{A2}
 V_{2}(n)=0.0528n^{4}+0.1n^{3}-0.836n^{2}+0.433n+0.0365. 
 \end{equation}
Values of both functions are in fm$^{-1}$.


\end{document}